\documentclass[aps,prl,reprint,superscriptaddress,nofootinbib]{revtex4-1}
\usepackage{graphicx}
\usepackage{epsfig}
\usepackage{rotate}
\usepackage{amsmath}
\usepackage{amssymb}
\usepackage{amsfonts}
\usepackage{enumerate}
\usepackage{afterpage}
\usepackage{xcolor}
\usepackage{bm}
\usepackage{hyperref}
\usepackage[T1]{fontenc}
\usepackage{ae,aecompl}
\usepackage{natbib}

\renewcommand{\mathbf}{\bm}
\newcommand{\n}{\bm {n}}
\renewcommand{\k}{\bm {k}}

\newcommand{\cH}{\mathcal{H}}

\def\be{\begin{equation}}
\def\ee{\end{equation}}
\def\bea{\begin{eqnarray}}
\def\eea{\end{eqnarray}}
\def\i{\mathrm{i\,}}

\newcommand{\mnras}{MNRAS}

\newcommand{\chris}[1]{{\color{black}{#1}}}
\newcommand{\pritha}[1]{{\color{black}{#1}}}

\newcommand{\response}[1]{{\color{black}{#1}}}
\newcommand{\newresponse}[1]{{\color{black}{#1}}}

\begin{document}
\title{\newresponse{Apparent} parity violation in the observed galaxy trispectrum}
\author{Pritha Paul}
\email{p.paul@qmul.ac.uk}
\affiliation{Astronomy Unit, School of Physical \& Chemical Sciences, Queen Mary University of London, London E1 4NS, UK}
\author{Chris Clarkson}
\affiliation{Astronomy Unit, School of Physical \& Chemical Sciences, Queen Mary University of London, London E1 4NS, UK}
\affiliation{Department of Physics \& Astronomy, University of the Western Cape, Cape Town 7535, South Africa}
\author{Roy Maartens}
\affiliation{Department of Physics \& Astronomy, University of the Western Cape, Cape Town 7535, South Africa}
\affiliation{National Institute for Theoretical and Computational Sciences, Cape Town 7535, South Africa}
\affiliation{ Institute of Cosmology \& Gravitation, University of Portsmouth, Portsmouth PO1 3FX, UK}



\date{\today}

\begin{abstract}
Recent measurements of the 4-point correlation function in large-scale galaxy surveys have found apparent evidence of parity violation in the distribution of galaxies. This cannot happen via {dynamical} gravitational effects in general relativity.  If such a violation arose from physics in the early Universe it could indicate important new physics beyond the standard model, and would be at odds with most models of inflation. It is therefore now timely to consider the galaxy trispectrum in more detail. {While} the \emph{intrinsic} 4-point correlation function, or equivalently the trispectrum, its Fourier counterpart, is parity invariant, the \emph{observed} trispectrum must take redshift-space distortions into account. Although the standard Newtonian correction also respects parity invariance, we show that sub-leading relativistic corrections do not. We demonstrate that these can be significant at intermediate linear scales and are dominant over the Newtonian parity-invariant part around the equality scale and above. Therefore when \emph{observing} the galaxy 4-point correlation function, we should expect to detect parity violation on large scales.

\end{abstract}
					
\maketitle
\paragraph{\bf\emph{Introduction}}

A detection of parity violation in large-scale galaxy statistics could be a signature of physics beyond the standard model.
Parity inversion, defined as reversing the sign {at any event} of each {spatial Cartesian} coordinate axis, is a symmetry that is obeyed by most physical processes. Amongst the fundamental forces, electromagnetic, strong interaction and gravitation are all invariant under parity transformations. 
However, parity violation occurs in the weak force and is a possible reason for baryon asymmetry among other things but how prevalent it is is not known \cite{Tokunaga_2013}.  
On cosmological scales, structure formation is dominated by gravitation which is parity invariant in the dominant scalar sector. Probing parity violation could provide an insight into early universe physics as we  expect it to be invariant under a parity transformation. 
In order to provide constraints on parity violation, we need to investigate observables that are sensitive to parity. Observables that are constructed from scalar fields, such as the galaxy density fluctuation, pose difficulties to being parity sensitive. For example, the power spectrum, the Fourier counterpart of the $2$ point correlation function (2PCF) is not sensitive to parity. 
With future galaxy surveys
we will constrain higher-order statistical measures such as the bispectrum and trispectrum. 
The trispectrum or $4$ PCF, is the lowest-order statistic that can probe parity violation in the scalar gravity sector. 

Several cosmological studies have considered the $4$PCF of the cosmic microwave background (CMB).
Most of the parity violation studies have been focused on the polarisation of the CMB \cite{Kamionkowski_2011,Minami_2020,PhysRevD.106.063503,Shiraishi_2011} or on gravitational waves \cite{Nishizawa_2018,Jeong_2020,Orlando_2021}. 
Recently, \cite{cahn2021test} have proposed using the galaxy $4$PCF as a test for detecting cosmological parity violation in large-scale structure, using quadruples of galaxies forming a tetrahedron, which is the lowest order $3$D shape which cannot be superimposed on its mirror image by a rotation. 
An imbalance between the tetrahedrons and their mirror images (assuming statistical isotropy) can provide information about parity violating parity-odd modes.
\cite{Philcox_2022} has done a blind test on the odd-parity part of the 4PCF using the Baryonic Oscillation Spectroscopic (BOSS) CMASS \citep{Rodr_guez_Torres_2016} sample of the Sloan Digital Sky Survey (SDSS) III \citep{Beutler_2014}. This test considered separations in the range $20-160 h^{-1} \mathrm{Mpc}$ and found $2.9 \sigma$ evidence for the detection of an odd-parity $4$PCF. 
\cite{Hou_2023} have also measured these parity-odd modes from BOSS,  performing an analysis of  DR$12$ LOWZ $(\Bar{z} =0.32)$ and  DR$12$ CMASS $(\Bar{z} =0.57)$ (\cite{slepian2015new}, \cite{2006MNRAS.372..425C}). They  found  $3.1\sigma$ and $7.1\sigma$ evidence for an odd-parity $4$PCF, respectively.

These findings imply that theoretical modelling of the galaxy trispectrum should now be investigated in more detail.
Galaxies are observed in redshift space on our past lightcone, and not in real space. The Kaiser redshift-space distortions (RSD) account for the first approximation to this \cite{Scoccimarro_1999}
on small scales, and these do not break the parity invariance of the trispectrum. But on larger scales, relativistic corrections come into play which are important \cite{McDonald:2009dh,Yoo_2010,Bonvin:2011bg,Bertacca:2012tp,Bonvin:2014owa,Bertacca:2014dra,Umeh:2016nuh,giusarma2017relativistic,Breton_2018}. 

We show that these relativistic contributions also break the parity invariance of the trispectrum. Effects such as peculiar velocities and gravitational redshift break the symmetry along the line of sight~-- real space spheres distort into redshift space egg shapes~-- and give rise to odd-multipoles in $n$-point statistics which break parity invariance \citep{Bonvin_2014,Clarkson:2018dwn,Jeong_20201}. \newresponse{This breaking of parity invariance is due to observer dependent effects arising from distortions of their past lightcone and is not intrinsic to the field. These effects alter the information that is available from the underlying dark matter distribution. However, these effects are present in the measurements as they are responsible for distorting the tetrahedral shapes that are probed by the trispectrum.} \response{As in \cite{Jeong_20201}, we refer to parity transformation as inverting the position of galaxies in an $n$-point statistic for a fixed line of sight}. \newresponse{That is, an observer would detect parity breaking in a patch of the sky, even though this is not intrinsic to the underlying galaxy distribution.}  These odd multipoles therefore reflect relativistic corrections which become important 
on larger scales. In the trispectrum we show that the odd parity-violating part picks up these important effects at order $\mathcal{H}/k$, \pritha{where $\mathcal{H}$ is the conformal Hubble rate}, which are surprisingly significant even on scales $k\sim 0.01-0.1h$\,Mpc$^{-1}$. While these corrections are not present at leading order in $\mathcal{H}/k$ in the monopole \response{(i.e., when $\mu$ is averaged over)} of the trispectrum (so are presumably unlikely to be the cause of the parity violation in \cite{Hou_2023}), they will play a role in future trispectrum analyses, in detecting large-scale relativistic effects.  

We now give a theoretical description of the observed galaxy trispectrum in a relativistic framework,  accounting for RSD, Doppler and gravitational redshift effects, up to third order in perturbation theory. 


\paragraph{\bf\emph{Parity violation in the power- and bi-spectra}}

The principal redshift-space effect on galaxy number counts at first order is given by the Kaiser RSD term, $\delta_g(\k) = (b_{1} + f\mu^{2})\delta(\k)$, where $\mu=\n\cdot\hat{\k}$, with $\n$ the line of sight direction, $f$ the growth rate, $b_1$  the linear bias (we omit the functional dependence on redshift here and below for convenience), \response{and $\delta(\k)$ the matter density contrast}. The principal local correction to this effect is a Doppler term~\cite{Kaiser:1987qv,McDonald:2009dh,Challinor:2011bk,Bertacca:2012tp} (see also \cite{Raccanelli:2016avd,Hall:2016bmm,Abramo:2017xnp,Jeong_2020})
proportional to $\bm{v}\cdot\n$, where $\bm{v}$ is the peculiar velocity:
\be \label{eq1}
\delta_g(\bm k)=\Big(b_1+f\mu^2+\i\mu {A}\,f {\cH}/{k} \Big)\delta(\k)=\mathcal{K}(\bm k)\delta(\k)\,,
\ee
where  ${A} = {b_{\rm e}+{3}\Omega_m/2 -3 + (2-5s) (1-{1/ r\cH} )}$.
We assume a $\Lambda$CDM background, implying $\cH'/\cH^2=1-3\Omega_m/2$, where 
$\Omega_m$ is the evolving matter density. The evolution of comoving galaxy number density is $b_{\rm e}=\partial {\ln} (a^3 \bar{n}_g)/\partial \ln a$ and $s=-(2/5)\partial \ln \bar{n}_g/\partial \ln L$ is  the magnification bias ($L$ is the threshold luminosity), while  $r$ is the comoving radial distance ($\partial_r=\bm n\cdot\bm\nabla$). 


The significance of each term can be estimated as a function of scale.
On small scales, terms 
$\sim\cH/k$ are suppressed, but they become more important near and above the equality scale. 
While $\delta_g$ in \eqref{eq1}
is no longer invariant under $\bm k\to-\bm k$ and 
is complex, 
the power spectrum is real and remains parity invariant 
since $\mu_{-\k}=-\mu_{\k}$.

The galaxy bispectrum requires the density contrast at second order. 
It is defined 
by
\begin{align} 
B_{g}
&= { \mathcal{K}( \k_{1})\mathcal{K}({\k}_{2}) \mathcal{K}^{(2)}(  \mathbf{k}_{1},  \mathbf{k}_{2}, \k_3)} 
P(k_{1})P(k_{2}) + 2 \, \text{perms} .\label{b6}
\end{align}
 $P$ is the matter power spectrum.
 The first-order kernel $\mathcal{K}=\mathcal{K}_{\rm N}+\mathcal{K}_{\rm R}$ is given by~\eqref{eq1}.
\chris{The second-order kernel has Newtonian and Relativistic parts,}  $\mathcal{K}^{(2)}=\mathcal{K}_{\rm N}^{(2)}+\mathcal{K}_{\rm R}^{(2)}$, where
 the Newtonian kernel is~\citep{Verde:1998zr,Tellarini:2016sgp} $\mathcal{K}_{\rm{N}}^{(2)}(\mathbf{k}_{1},  \mathbf{k}_{2},\k_3)
 = b_{2}+ b_{1}F_{2} + f\mu_3^{2}G_{2} + {\cal Z}_2$
where $F_2({\k_{1}},  {\k_{2}}),G_2({\k_{1}},  {\k_{2}})$ are the second-order density and velocity kernels, and ${\cal Z}_2({\k_{1}},  {\k_{2}})$ is the second-order RSD kernel \chris{and $b_2$ is the second-order bias}. We use a local bias model \citep{Desjacques:2016bnm}, but ignore tidal bias \chris{for simplicity}. 

The leading relativistic correction at second order is 
\cite{Bertacca:2014dra,Bertacca:2014hwa,Yoo:2014sfa,DiDio:2014lka,Jolicoeur:2017nyt,Dio_2019}:  
\begin{align} 
&
\delta^{(2)}_{g{\rm R}}=A\, \bm{v}^{(2)}\!\!\cdot\n+2{C}(\bm{v}\cdot\n)\delta +2{{E}\over\cH}(\bm{v}\cdot\n)\partial_r(\bm{v}\cdot\n)
\\&~\nonumber
+2{b_1\over\cH}\phi\, \partial_r\delta
+{2\over\cH^2}\big[\bm{v}\cdot\n \,\partial_r^2\phi-\phi\, \partial_r^2 (\bm{v}\cdot\n) \big] -{2\over \cH}\partial_r (\bm{v}\cdot\bm{v}),  \label{dg2}
\end{align}
where \pritha{$\boldsymbol{v} \cdot \boldsymbol{n}$ is the peculiar velocity along the line of sight}, $\phi$ is the gravitational potential,   $C = b_1(A + f)+{b_1'/ \cH}+ 2(1-{1/ r\cH} ){\partial b_1/ \partial\ln L}$ and
$E = {4-2A-3\Omega_m/2}$. 
The corrections in $\delta_g^{(2)}$ all scale as $(\cH/k)\delta^2$ and so can be significant on scales approaching equality and above. 
These corrections appear in the bispectrum in the kernel~\citep{Jolicoeur:2017eyi},
\begin{align}
&\mathcal{K}^{(2)}_{\mathrm{R}}
=  {\i} {\cH}
\Big[-{3}\mu_{1}{k_{1}\over k_2^2}\Omega_m b_1 
+4\nu_{12} {\mu_{1}\over k_2} f^2 
+ 2{\mu_{1}\over k_1} {C} f 
 \\&
-{3} \mu_{1}^3{k_{1}\over k_2^2} \Omega_mf 
 + {\mu_{1}^2\mu_{2}\over k_2} {\Big(3\Omega_m-2E f\Big) f}
+ {\mu_{3} \over k_{3}}G_{2}
{A} f\Big]_{\circlearrowright_{1,2}} 
\nonumber
\end{align}
where $\nu_{ij}=\hat{\k}_i\cdot \hat{\k}_j$ and $\mu_i=\hat{\k}_i\cdot\bm n$. Here $\circlearrowright$ denotes symmeterisation\footnote{We use this in the tensor notation manner, i.e., $$
X(\bm k_1\cdots\bm k_n)_{ {\circlearrowright}_{\tiny{1\cdots n}}}=\frac{1}{n!}\left[X(\bm k_1\cdots\bm k_n)+\text{all perms of~}\bm k_1\cdots\bm k_n\right]
$$
} on $\bm{k}_{1\cdots2}$
The Newtonian kernel 
scales as $(\cH/k)^{0}$, while the Doppler kernel 
scales as $(\cH/k)$. More generally there are many other corrections to the relativistic kernel but these scale as $(\cH/k)^2$ or higher, and are sub-dominant~\cite{deWeerd:2019cae}.

The bispectrum will be parity invariant provided it is invariant under a sign change of each $\bm{k}_i$. 
The parity-violating part of the bispectrum is found by calculating 
\begin{align}
B_{\text{odd}}
&=\frac{1}{2}\Big[B_{g}( \mathbf{k}_{1},  \mathbf{k}_{2},  \k_3)
- B_{g}( -\mathbf{k}_{1},  -\mathbf{k}_{2},  -\k_3) \Big]\,. 
\end{align}
First, under $\bm{k}_i\to-\bm{k}_i$ we have $ k_i\to k_i$ and $\mu_i\to-\mu_i$.  Contributions with odd powers of $\mu_i$ will not be parity invariant. The Newtonian kernels consist of terms with even powers of $\mu_i$ only (including zero), while leading-order relativistic terms consist only of odd powers of $\mu_i$. To order $(\cH/k)$ which we consider here, in forming the bispectrum by multiplying out the kernels $\mathcal{K}_N+\mathcal{K}_R$ we see that the bispectrum structure is
Newtonian terms $\mathcal{O}(\cH/k)^0\times\,$even powers of $\mu_i$'s +
{Relativistic terms} $\i\times\mathcal{O}(\cH/k)^1\times\,$ {odd powers of} $\mu_i$'s.
Then $B_{\text{odd}}\sim $ {Relativistic terms} $\i\times\mathcal{O}(\cH/k)^1\times\,${odd powers of} $\mu_i$'s,
so that relativistic terms introduce parity violation in the observed bispectrum. This  is equivalent to the bispectrum having odd multipoles~\citep{deWeerd:2019cae,Jeong_20201}.

\paragraph{\bf\emph{Relativistic contributions to the trispectrum}}


In the trispectrum, relativistic terms also give rise to an imaginary part which changes sign under a parity transformation. 
The imaginary part of the trispectrum is a direct consequence of relativistic effects, which are not present in a Newtonian analysis. 
In order to compute this trispectrum, it requires us to go up to third order in perturbation theory.
At third order, the RSD and Doppler corrections generalise to $\bm v^{(3)}\cdot\bm n$ 
together with a large variety of coupling terms between different orders. 
These terms include radial gradients of the potential and the density contrast, as well as a host of Doppler contributions at each order, including transverse velocities. 
The leading correction with all the local relativistic effects up to order $\mathcal{H}/k$ was calculated in \cite{Dio_2019}\footnote{Our $\boldsymbol{n}$ is minus theirs;  their convention $\delta_g^{(1)}  + \delta_g^{(2)}+ \delta_g^{(3)}$ differs from ours: $\delta_g^{(1)}  + \delta_g^{(2)}/2+ \delta_g^{(3)}/3!$. We alter the standard kernels $F_2, G_2, F_3, G_3$ accordingly.}. 
 All these relativistic corrections are a consequence of the projection along the line of sight $\boldsymbol{n}$ in redshift space\footnote{We neglect terms that are integrated along the line of sight, such as lensing magnification and the integrated Sachs-Wolf effect. We adopt the standard plane-parallel approximation. 
 We drop any terms that scale as $(\mathcal{H}/{k})^2$ and higher, since these do not contribute to the parity violation that we  consider. 
 We also do not include any primordial non-Gaussianity.}.
The trispectrum of the observed galaxy number density fluctuation is defined by
\begin{align}
    \big\langle \delta_g(\boldsymbol{k}_1) \delta_g(\boldsymbol{k}_2) \delta_g(\boldsymbol{k}_3) &\delta_g(\boldsymbol{k}_4) \big\rangle = (2\pi)^3 T_{g}(\boldsymbol{k}_1,\boldsymbol{k}_2,\boldsymbol{k}_3, \bm k_4) \notag \\
    &~~~ \times \delta^D(\boldsymbol{k}_1+\boldsymbol{k}_2+\boldsymbol{k}_3+\boldsymbol{k}_4). 
\end{align}
At third order, the only combination of terms that contribute at tree level are
\begin{align}
    &\big\langle \delta_g(\boldsymbol{k}_1) \delta_g(\boldsymbol{k}_2) \delta_g(\boldsymbol{k}_3) \delta_g(\boldsymbol{k}_4)\big\rangle = 
    \\
    &\qquad\qquad~~
    \Big[4\langle \delta_g^{(1)}(\boldsymbol{k}_1) \delta_g^{(1)}(\boldsymbol{k}_2)\delta_g^{(1)}(\boldsymbol{k}_3) \delta_g^{(3)}(\boldsymbol{k}_4)\rangle
    \notag \\
    &\qquad\qquad~~~~ 
    +6 \langle \delta_g^{(1)}(\boldsymbol{k}_1) \delta_g^{(1)}(\boldsymbol{k}_2) \delta_g^{(2)}(\boldsymbol{k}_3) 
    \delta_g^{(2)}(\boldsymbol{k}_4)\rangle \Big]_{\circlearrowright_{1\cdots4}} \notag.
\end{align}
Symmetrisation gives \pritha{4} distinct permutations of type 1113 and \pritha{6} of type 1122.\footnote{Contributions of the form 1111 are just power spectra squared and removed from the analysis, and contributions 1112 are zero under the assumption of Gaussian initial conditions.}  
Using Wick's theorem, this leads to 
\begin{align}
&\pritha{T_g}
= \notag
\Big\{6 \mathcal{K}^{(3)}(-\boldsymbol{k}_1, -\boldsymbol{k}_2, -\boldsymbol{k}_3, \boldsymbol{k}_4) 
\mathcal{K}^{(1)}(\boldsymbol{k}_1)\mathcal{K}^{(1)}(\boldsymbol{k}_2) \mathcal{K}^{(1)}(\boldsymbol{k}_3)
\\&\times  P(k_1) P(k_2) P(k_3)\Big\} + 3 \,\mathrm{cyc} \,\mathrm{perm}
\notag   
\\ \notag    &~
    +\Big\{4 \mathcal{K}^{(1)}(\boldsymbol{k}_1) \mathcal{K}^{(1)}(\boldsymbol{k}_2) \mathcal{K}^{(2)}(-\boldsymbol{k}_1, \boldsymbol{k}_1+\boldsymbol{k}_3, \boldsymbol{k}_3)
    \notag 
    \\\notag &
    \times
    \mathcal{K}^{(2)}(-\boldsymbol{k}_2, -\boldsymbol{k}_1\!-\boldsymbol{k}_3, \boldsymbol{k}_4)  P(|\boldsymbol{k}_1 + \boldsymbol{k}_3|) 
    \\\notag &
    + 4 \mathcal{K}^{(1)}(\boldsymbol{k}_1) \mathcal{K}^{(1)}(\boldsymbol{k}_2) \mathcal{K}^{(2)}(-\boldsymbol{k}_2, \boldsymbol{k}_2+\boldsymbol{k}_3, \boldsymbol{k}_3)
    \notag 
    \\\notag &
    \times
    \mathcal{K}^{(2)}(-\boldsymbol{k}_1, -\boldsymbol{k}_2\!-\boldsymbol{k}_3, \boldsymbol{k}_4)  P(|\boldsymbol{k}_2 + \boldsymbol{k}_3|)\!\Big\}P(k_1) P(k_2)
    \\ &
    + 5 \,\mathrm{cyc} \,\mathrm{perm} 
\end{align}
We  break down the third-order kernel into  Newtonian and relativistic parts,  
$\mathcal{K}^{(3)} = \mathcal{K}^{(3)}_{\mathrm{N}} + \mathcal{K}^{(3)}_{\mathrm{R}}.$
The Newtonian kernel is \citep{Bernardeau_2002,Philcox_2022}:
\begin{align} \label{newtkern}
    &\pritha{\mathcal{K}^{(3)}_N} = \Big[b_1F_3 + 3b_2F_2 + b_3 + \mu_4^2fG_3  
    \notag \\
    &+ \frac{3\mu_3}{k_3}(\mu_1 k_1 + \mu_2 k_2 + \mu_3 k_3 )f\Big(b_1F_2 + b_2 
    \notag\\
    &+ \frac{(\mu_1 k_1 +\mu_2k_2)^2}{|\boldsymbol{k}_1 + \boldsymbol{k}_2|^2}fG_2 \Big) + \frac{3(\mu_1 k_1 +\mu_2k_2)}{|\boldsymbol{k}_1 + \boldsymbol{k}_2|^2} \notag \\
    &(\mu_1 k_1 +\mu_2k_2 + \mu_3 k_3) f G_2(b_1 + \mu_3^2f) \notag \\
    &+ \frac{\mu_2 \mu_3}{ k_2 k_3}(\mu_1 k_1 + \mu_2 k_2 + \mu_3 k_3)^2 f^2 \notag\\
    &\Big(\frac{\mu_1}{k_1}(\mu_1 k_1 + \mu_2 k_2 + \mu_3 k_3)f + 3b_1 \Big)\Big]\pritha{_{\circlearrowright_{1\cdots3}}}
\end{align}
Here $F_3(\boldsymbol{k}_1, \boldsymbol{k}_2, \boldsymbol{k}_3)$ and $G_3(\boldsymbol{k}_1, \boldsymbol{k}_2, \boldsymbol{k}_3)$ are the third order density and velocity kernels adapted to our notation \citep{Jain_1994}. 
The relativistic correction to \eqref{newtkern} in Fourier space follows from \cite{Dio_2019} and the full expression is given in the Appendix (a follow-up paper will present the detailed derivation). 
We have verified numerically that the third-order relativistic kernel is required for an accurate calculation of the trispectrum, as the 1113 contribution is similar to the 1122, depending on the configuration. \response{For example, in certain configurations, the 1113 dominates 1122, in others it cancels part of the 1122 contribution~-- see the Appendix.}

\paragraph{\bf\emph{Parity violation in the trispectrum}}

The parity-preserving and parity-violating parts of the trispectrum can be found by calculating 
\begin{align}
T_{\substack{\text{even} \\ \text{odd}}}
    &= \frac{1}{2} \Big[T_g(\boldsymbol{k}_1, \boldsymbol{k}_2, \boldsymbol{k}_3, \boldsymbol{k}_4)
\pm T_g(-\boldsymbol{k}_1, -\boldsymbol{k}_2, -\boldsymbol{k}_3, -\boldsymbol{k}_4)\Big]\!.
\end{align}
We expect $T_\text{odd}\neq0$ for the same reason that the observed bispectrum is parity violating. The terms in the relativistic kernels have only odd powers of $\boldsymbol{n}$ and odd powers of $\mu_i$.\footnote{By this we mean that any term with $\mu_1^{i_1}\mu_2^{i_2}\mu_3^{i_3}\mu_4^{i_4}$ has $i_1+i_2+i_3+i_4$ an even number or zero.}  The terms responsible for the odd powers in the relativistic kernels include contributions such as $\bm v\cdot\bm n$ and $\partial_r\delta$. As a result, the terms that scale as $(\mathcal{H}/k)^1$ will introduce parity violation in the trispectrum, while the terms that scale as $(\mathcal{H}/k)^0$ and even powers of $\mu_i$ will not. 
We calculate the even and odd parts neglecting terms $(\mathcal{H}/k)^2$ and higher.
This ensures that the even part here is purely Newtonian, $O(\mathcal{H}/k)^0$, while the odd part will arise from a coupling of the leading relativistic part with Newtonian terms. More generally, relativistic terms would enter the even part at order $(\mathcal{H}/k)^2$ (the same order as local primordial non-Gaussianity at the perturbative order that we are considering), \response{and we expect a hierarchy of contributions at many powers of $(\i\mathcal{H}/k)$. In general, we can expand $T_g$ in spherical harmonics about $\bm n$, with even multipoles containing even powers of $(\i\mathcal{H}/k)$ (making up the real part), while odd multipoles contain odd powers of $(\i\mathcal{H}/k)$, making up the odd imaginary part of $T_g$~\cite{deWeerd:2019cae}.}

We now turn to calculating the amplitude of the odd parity part relative to the even Newtonian part. 
We let $\bm k=\bm k_1+\bm k_2 = -\bm k_3-\bm k_4$ be the vector which breaks the quadrilateral into two triangles. We choose the $z$ axis in the direction of $\bm k$ and place the triangle $\bm k-\bm k_1-\bm k_2=0$ in the $x=0$ plane. Let \response{$\pi-\Theta$} be the angle between the $z$-axis and $\bm k_2$ and $\Phi$ be the angle between the $z$-axis and $\bm k_3$. The angle between the 2 quadrilateral triangle flaps is $\Psi$.  Then the shape is fixed by the set 
$k, s=k_2/k, t=k_3/k, \Theta, \Phi, \Psi$ implying
 $k_1=k\sqrt{1+s^2+2s\cos\Theta}\,,~~ k_4=k\sqrt{1+t^2+2t\cos\Phi}\,.$   
Finally $\bm n$ is an arbitrary vector with spherical coordinates $(\theta,\phi)$. \response{(A figure of the setup is given in the Appendix.)}

To demonstrate the amplitude of the odd-parity modes, we consider two types of survey which will be affected by the relativistic terms via the magnification and evolution biases. The first survey is an SKA-like intensity mapping (IM) of 21\,cm neutral hydrogen radio emission  at $z=1$. This has bias parameters 
$b_1 = 0.856, b_2 = -0.321, b_1' = -0.5\times10^{-4} \response{\mathrm{Mpc^{-1}}}, b_e = -0.5, b_e'=0, s = 2/5$~\citep{Umeh:2015gza,Fonseca:2018hsu} (we ignore tidal bias and set third-order bias to zero for simplicity). Second is a Euclid-like \response{near-}infrared spectroscopic survey, with 
$ b_1 = 1.3,b_2 = -0.74, b_1' = -1.6\times10^{-4} \response{\mathrm{Mpc}^{-1}},  b_e = -4, b_e' = 0, s = -0.95$~\citep{Camera:2018jys,Yankelevich:2018uaz}.
(We set $ \partial b_1/\partial \ln L =0$ for simplicity.) The $\Lambda$CDM model has parameters $\Omega_{m0}=0.31, h=0.68, f_\text{baryon}=0.157, n_s=0.968$. Plots are presented using linear power spectra with the fitting formula of~\cite{Eisenstein:1997ik}.

We consider the behaviour as a function of scale, viewing orientation, and tetrahedron shape. 
\begin{figure}
    \centering
\includegraphics[width=0.95\linewidth]{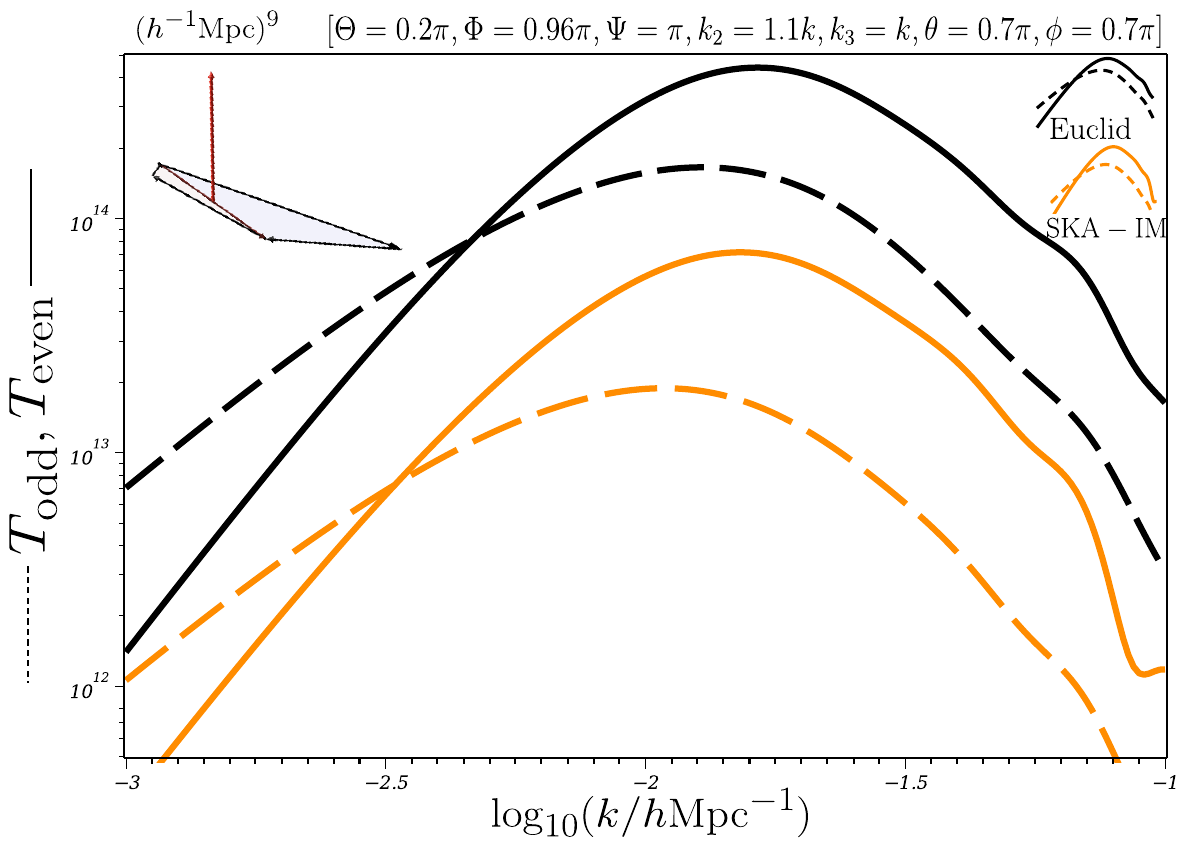}
    \caption{The odd (dashed) and even (solid) trispectra as a function of scale, for both Euclid-like and SKA-like surveys. The geometry -- flat quadrilateral with the line of sight along $\bm k$ --  is shown on the inset. In both cases, the odd part is dominant on large scales, and important even on intermediate scales.}
    \label{fig:IM-Euclid}
\end{figure}
Figure~\ref{fig:IM-Euclid} displays the odd and even parts of the trispectrum for both Euclid- and SKA-like surveys at $z=1$. These are for a flattened configuration with the tetrahedron lying in a plane ($\Psi=\pi$), viewed along the folding vector $\bm k$ (the $z$-axis in our coordinates). 
\response{Looking for example at the Euclid case (black curves), the even part peaks at $\sim 0.02 h$Mpc$^{-1}$, while on larger scales, $\sim 0.017 h$Mpc$^{-1}$ the odd part peaks, and is nearly as large as the even part at that scale and becomes larger above (i.e., for smaller $k$).}
We find in this example that 
corrections 
 for the \response{Euclid and SKA} like case are larger than 10\% \response{(i.e., $T_\text{odd}>0.1T_\text{even}$)} \response{for all scales}. The parity-violating part is similar in size to the parity-preserving part for $k\lesssim 0.006h$\,Mpc$^{-1}$ in the Euclid-like case and
$k\lesssim 0.0\response{05}h$\,Mpc$^{-1}$ for SKA-like. 


\begin{figure}
    \centering
    \includegraphics[width=0.99\linewidth]{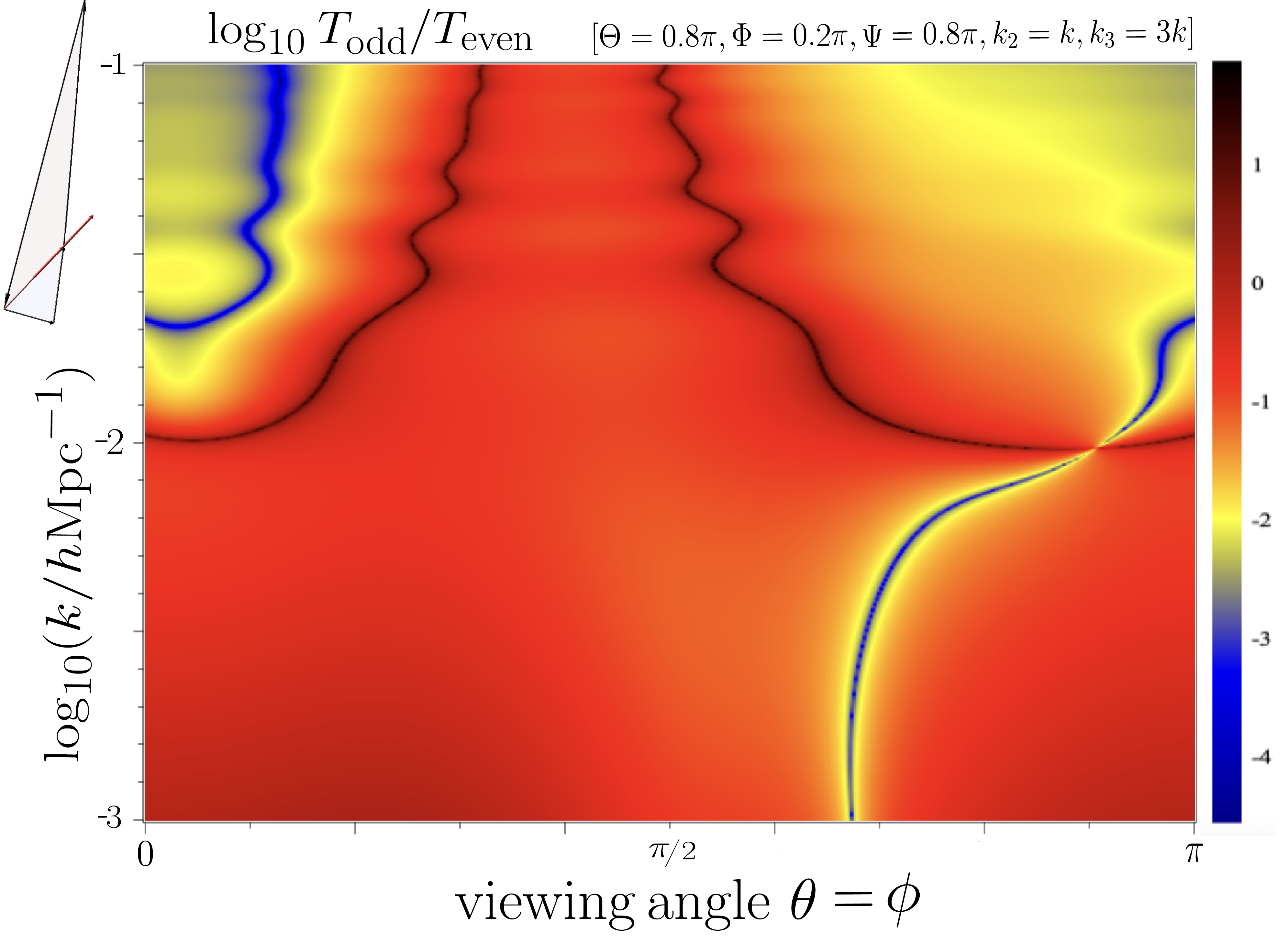}
    \caption{The ratio of the odd to even trispectrum for a \response{Euclid}-like survey as a function of scale and viewing orientation, \response{with  $\theta=\phi$}. Note the colouring is a log scale, and the dark red curves are where the even part changes sign. }
    \label{fig:IM1}
\end{figure}

In Fig.~\ref{fig:IM1} we show the ratio of the odd- to even-parity trispectrum for an Euclid-like survey at $z=1$. We fix the shape of the quadrilateral and change the scale and viewing orientation as indicated~-- \response{i.e., as we spin the tetrahedron around.} \response{The colouring on a log scaling indicates the amplitude of the corrections, and, roughly speaking, regions with colouring of orange or red indicates corrections of 10\% or more.} We see that on all scales $k\lesssim 0.1h$\,Mpc$^{-1}$, the parity-violating part is substantial, well over 10\% -- and is clearly dominant above the equality scale, \response{i.e., on the lower half of the plot. The Baryon Acoustic Oscillation are clearly visible below the equality scales for some viewing angles.} 


\paragraph{\bf\emph{Conclusions}}

We have investigated for the first time the effect that large-scale relativistic RSD have on the observed galaxy trispectrum. These relativistic projection effects induce parity violation even in the case of a primordial trispectrum that is parity preserving. The effects that need to be included are considerably more complicated than in the Newtonian RSD case, and a third-order analysis is required. The leading relativistic contribution is of order $\cH/k$ and  changes sign under $\bm k\to-\bm k$, thus breaking the parity invariance of the Newtonian contribution. 
We have shown that the changes are significant, and cannot necessarily be neglected, even for scales $k\sim0.1h$\,Mpc$^{-1}$, with corrections around 5\%, becoming even more significant on larger scales~-- and dominant on equality scales and above. Although the monopole does not have this parity violation as a result of the relativistic contributions, the fact that the dipole has such a significant contribution implies that measurements of the monopole have to be very careful to remove possible contaminants \response{in the observed trispectrum} which are naturally parity-violating. \response{In the case of parity violation that has tentatively been observed~\cite{Philcox_2022}, it may be unlikely that relativistic effects have caused this because they peak on larger scales, but there could be some contamination from the intrinsic dipole coupling with the window function.} \response{Nevertheless, the results presented here generalise and are similar in relative size to those found for the galaxy bispectrum -- as these are expected to be detectable with upcoming surveys~\cite{Maartens:2019yhx,Jolicoeur:2020eup} we anticipate the odd part of the trispectrum should be similarly detectable. }

\paragraph{\bf\emph{Acknowledgements}} We thank Dionysis Karagiannis and Scott Melville for discussions. 
RM is supported by the South African Radio Astronomy Observatory and National Research Foundation (Grant No. 75415). 

\vspace{2cm}


\begin{widetext}

\appendix

\section{ Third-Order Relativistic Kernel } \label{app1}

\begin{equation}
\begin{split}
\mathcal{K}^{(3)}_{\mathrm{GR}} &=   \mathrm{i} f \mathcal{H} \Bigg \{ - \Bigg(\frac{{\mathcal{\dot{H}}}}{\mathcal{H}^2} + \frac{2 - 5s}{\mathcal{H}r} + 5s - b_e \Bigg) \Bigg[\frac{\mu_4}{k_4}  G_3(\boldsymbol{k}_1, \boldsymbol{k}_2, \boldsymbol{k}_3) +  \frac{3\mu_3}{k_3} \big[b_1 F_2(\boldsymbol{k}_1 , \boldsymbol{k}_2) +  b_2\big] +  \frac{3(\mu_1k_1 + \mu_2 k_2)}{|\boldsymbol{k}_1 + \boldsymbol{k}_2|^2} b_1G_2(\boldsymbol{k}_1, \boldsymbol{k}_2) \Bigg]
\\
&-3f G_2(\boldsymbol{k}_1, \boldsymbol{k}_2) \Bigg(1 +  \frac{3\dot{\mathcal{H}}}{\mathcal{H}^2} + \frac{4-5s}{\mathcal{H}r} + 5s - 2 b_e \Bigg)\Bigg[ \frac{\mu_3}{k_3}\Bigg(\frac{(\mu_1k_1 + \mu_2 k_2)}{|\boldsymbol{k}_1+ \boldsymbol{k}_2|}\Bigg)^2  + \frac{\mu^2_3(\mu_1k_1 + \mu_2 k_2)}{|\boldsymbol{k}_1+ \boldsymbol{k}_2|^2} \Bigg] 
\\
& + 3f \Bigg( \frac{\mu_3}{k_3} \Big[ b_1 F_2(\boldsymbol{k}_1 , \boldsymbol{k}_2) +  b_2\Big] +  \frac{(\mu_1k_1 + \mu_2 k_2)}{|\boldsymbol{k}_1 + \boldsymbol{k}_2|^2}  b_1 G_2(\boldsymbol{k}_1, \boldsymbol{k}_2)  \Bigg)
\\
&+ 6 f  G_2(\boldsymbol{k}_1, \boldsymbol{k}_2) \Bigg[\frac{(\mu_1k_1 + \mu_2 k_2)  (\boldsymbol{k}_3 \cdot (\boldsymbol{k}_1 + \boldsymbol{k}_2) - \mu_3 k_3 (\mu_1k_1 + \mu_2 k_2))}{|\boldsymbol{k}_1 + \boldsymbol{k}_2|^2k_3^2}    + \frac{\mu_3\Big\{(\boldsymbol{k}_1 + \boldsymbol{k}_2) \cdot \boldsymbol{k}_3 - (\mu_1k_1 + \mu_2 k_2)\mu_3k_3\Big\}}{|\boldsymbol{k}_1 + \boldsymbol{k}_2|^2k_3}  \Bigg] 
\\
& - \frac{9}{2} \Omega_m \Bigg\{ \Bigg[ \frac{(\mu_1 k_1 + \mu_2 k_2)^3}{|\boldsymbol{k}_1 + \boldsymbol{k}_2|^2k_3^2} G_2(\boldsymbol{k_}1, \boldsymbol{k}_2) +  \frac{\mu_3^3 k_3}{|\boldsymbol{k}_1 + \boldsymbol{k}_2|^2} F_2(\boldsymbol{k}_1, \boldsymbol{k}_2) \Bigg]  + \frac{1}{f} \Bigg\{ \frac{(\mu_1 k_1 + \mu_2 k_2)}{k_3^2} \Big[b_1 F_2(\boldsymbol{k}_1, \boldsymbol{k}_2) +  b_2\Big]  \\
&+ \frac{\mu_3 k_3}{|\boldsymbol{k}_1 + \boldsymbol{k}_2|^2 }  b_1 F_2(\boldsymbol{k}_1, \boldsymbol{k}_2)\Bigg\}  -   \Bigg[ \frac{\mu_3^2(\mu_1 k_1 + \mu_2 k_2)}{|\boldsymbol{k}_1 + \boldsymbol{k}_2|^2}  G_2(\boldsymbol{k}_1, \boldsymbol{k}_2)   +  \frac{\mu_3 (\mu_1 k_1 + \mu_2 k_2)^2}{|\boldsymbol{k}_1 + \boldsymbol{k}_2|^2k_3}  F_2(\boldsymbol{k}_1, \boldsymbol{k}_2)  \Bigg] \Bigg\} \\
&- 6\Bigg(\frac{9}{2} \Omega_m f \mu_1^2 \mu_2 \Bigg(\frac{\mu_2^2 k_2}{k_3^2}  - \frac{\mu_3^2}{k_2}  \Bigg)  - \frac{3}{2} \Omega_m \Bigg\{f\frac{\mu_1}{k_1} \Bigg(\frac{ \mu_2^4 k_2^2}{ k_3^2}  - \frac{ \mu_2 \mu_3^3 k_3}{ k_2}   \Bigg) -  \frac{\mu_2}{k_2}   \Bigg[ \frac{(\mu_2 k_2 + \mu_3 k_3)^2}{k_1^2 }   - \mu_1^2  b_1  \Bigg] \Bigg\} 
\\
&- \frac{\mu_1 \mu_2 \mu_3}{3k_1k_2k_3}f^2 \Bigg(1 + \frac{3{\mathcal{H}'}}{\mathcal{H}^2} + \frac{3}{\mathcal{H}r} - \frac{3}{2} b_e \Bigg)  \Big(\mu_1 k_1 + \mu_2 k_2 + \mu_3 k_3\Big)^2  
\\
&- \frac{\mu_1 \mu_2}{2 k_1 k_2} f \Bigg(1 + \frac{3{\mathcal{H}'}}{\mathcal{H}^2} + \frac{4}{\mathcal{H}r} - 2 b_e \Bigg) \Big(\mu_1k_1 + \mu_2 k_2 + \mu_3 k_3\Big)  - \frac{\mu_1 \mu_2}{k_1 k_2}f^2 \Big(\mu_1k_1 + \mu_2 k_2 + \mu_3 k_3\Big) 
\\
&+ \big(\boldsymbol{k}_2 \cdot \boldsymbol{k}_3 - \mu_2 k_2 \mu_3 k_3\big) \Bigg[  \frac{f^2\big(\mu_1 k_1 + \mu_2 k_2 + \mu_3 k_3\big)^2}{2k_1 k_2^2 k_3^2}  +  \frac{f\big(\mu_1 k_1 + \mu_2 k_2 + \mu_3 k_3\big)}{k_2^2 k_3^2} \Bigg]
\\
& - \frac{f}{2}  \frac{ \mu_3 k_3}{k_1^2 k_2^2} (\boldsymbol{k}_2 \cdot \boldsymbol{k}_3)  + 2f^2 \frac{ \mu_1 \mu_3}{k_1 k_3}   \Bigg[\frac{ (\mu_1 k_1 + \mu_2 k_2 + \mu_3 k_3)}{k_2^2}\big(\boldsymbol{k}_1 \cdot \boldsymbol{k}_2 - \mu_1 k_1 \mu_2 k_2\big) \Bigg] \Bigg) \Bigg\} _\pritha{_{\circlearrowright_{1\cdots3}}}.
\end{split}
\end{equation}

\end{widetext}

\appendix

\section{Second and third order contributions} \label{app3}

\begin{figure}
\includegraphics[width=1\linewidth]{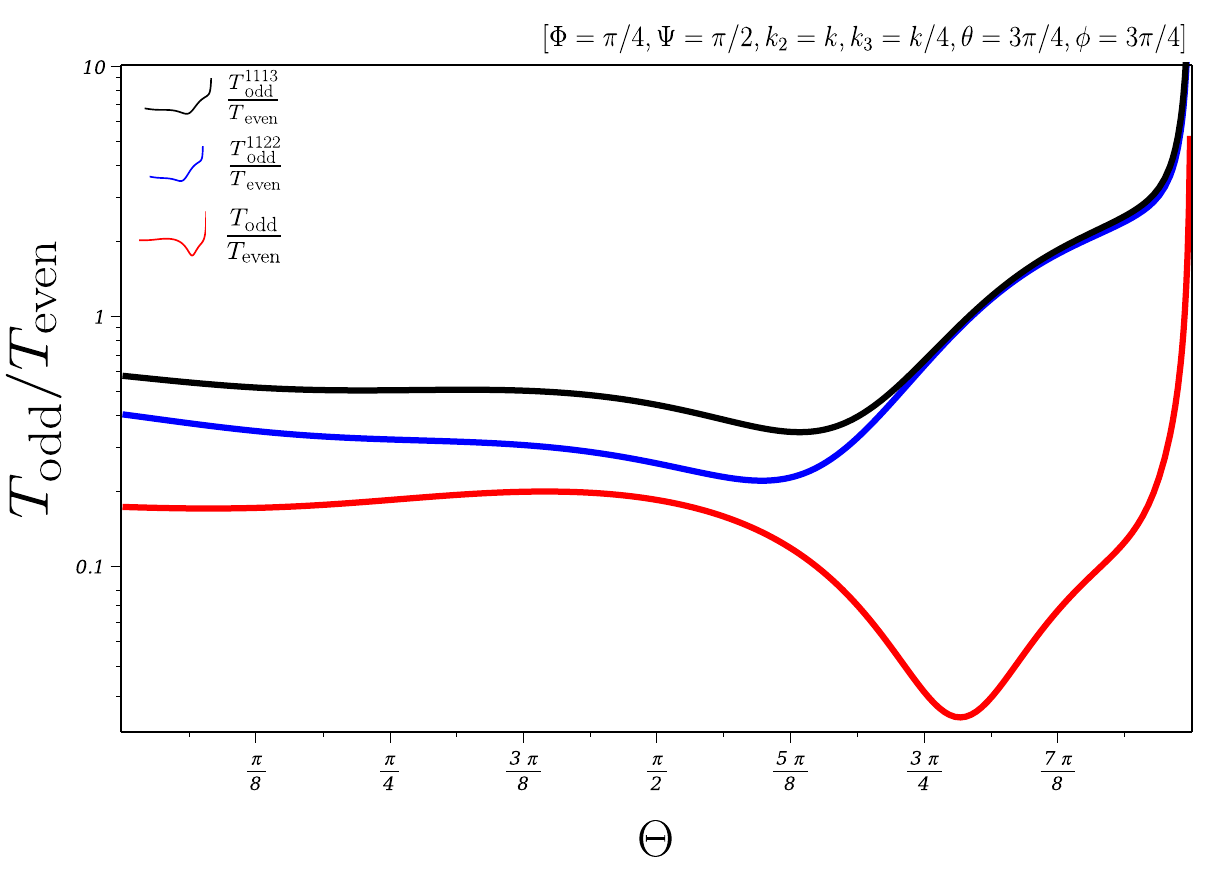}
\caption{The ratio of the $1113$ and $1122$ odd part to the even part of the trispectrum for a Euclid-like survey as a function of angle $\Theta$ at $k = 0.01 h \mathrm{Mpc}^{-1}$. Note that we have presented the absolute values of $T_{\mathrm{odd}}$ and $T_{\mathrm{even}}$ and the total signal is usually smaller than the individual 1122 and 1113 contributions indicating partial cancellation between some of these terms.}
\label{fig:varyTheta}
\end{figure}

We define $T_{1113}$ and $T_{1122}$ as the third and second order contributions to the trispectrum respectively. 
\newresponse{Figure \ref{fig:varyTheta} displays the ratios varying with $\Theta$. For this particular configuration, we see that the $T_{\mathrm{odd}}^{1113}/T_{\mathrm{even}}$ dominates for values of $\Theta \lesssim 3\pi/4$. For values of $0.75 \pi \lesssim \Theta \lesssim 0.86 \pi$, the contributions $T_{\mathrm{odd}}^{1113}/T_{\mathrm{even}}$ and $T_{\mathrm{odd}}^{1122}/T_{\mathrm{even}}$ are close to each other. The third order and second order contributions dominates for different values of the orientation and configuration angles. }

\begin{figure}

\includegraphics[width=1\linewidth]{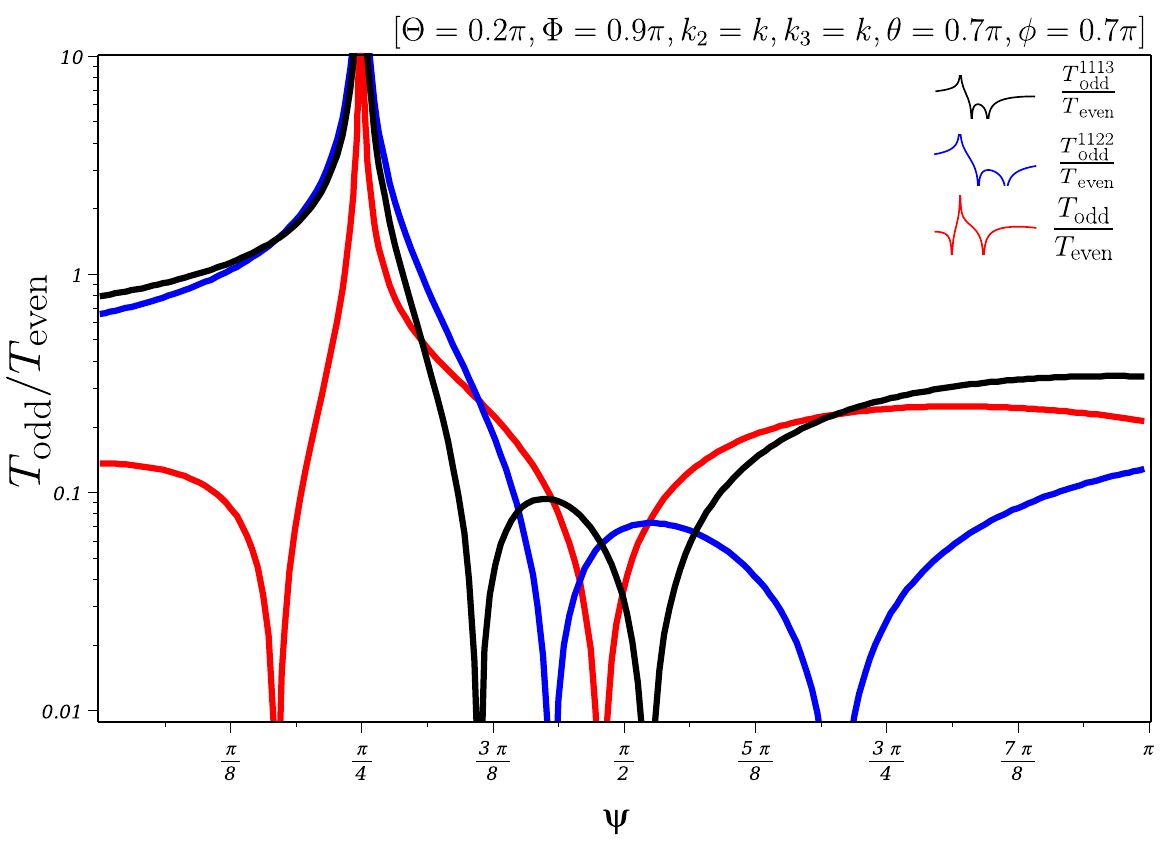}
\caption{The ratio of the $1113$ and $1122$ odd part to the even part of the trispectrum for a Euclid-like survey as a function of angle $\Psi$ at $k = 0.01 h \mathrm{Mpc}^{-1}$. Note that we have presented the absolute values of $T_{\mathrm{odd}}$ and $T_{\mathrm{even}}$ and the total signal varies according to the addition and cancellation of some of the terms in the individual 1122 and 1113 contributions.}
\label{fig:varypsi}
\end{figure}

Figure \ref{fig:varypsi} displays the ratios varying with the folding angle $\Psi$. Flattened configuration refers to the angle $\Psi$ being equal to $\pi$, where the two triangles lie on the same plane. It  shows that the ratio $T_\mathrm{\mathrm{odd}}^{1113}/T_{\mathrm{even}}$ and $T_{\mathrm{odd}}^{1122}/T_{\mathrm{even}}$ significantly changes as we vary the geometry of the quadrilateral. \newresponse{From the graph, we see the trend that $T_{\mathrm{odd}}^{1113}/T_{\mathrm{even}}$ dominates for most values of $\Psi$ except when $\pi/6 \lesssim \Psi \lesssim 3\pi/7$. At $\Psi \approx \pi/4$, we have $T_{\mathrm{even}} = 0$, and all three contributions tend to infinity. At $\Psi \approx 0.47 \pi $ and $0.17 \pi$, $T_{1113}$ and $T_{1122}$ changes signs and are exactly equal to each other. The total contribution is $0$ at these points. There are also interesting cancellations happening between the two sets of contributions. We will be investigating these in more details in a follow-up paper.}

\begin{figure} [h!]
\includegraphics[width=1\linewidth]{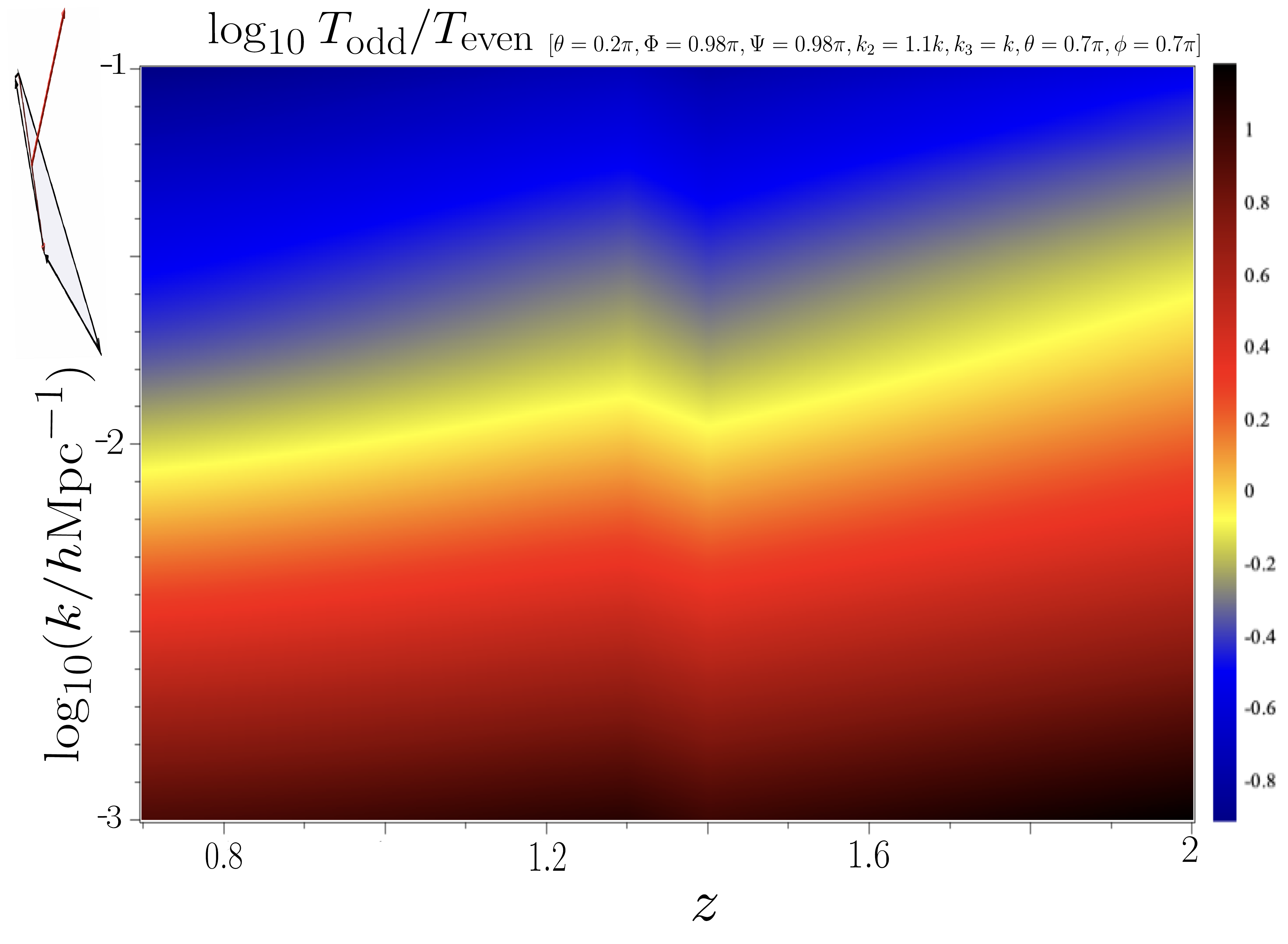}
\caption{The ratio of the odd to even trispectrum for a \response{Euclid}-like survey as a function of scale and redshift. Note the colouring is a log scale.}
\label{fig:Tvaryredshift}
\end{figure}

\newresponse{Figure \ref{fig:Tvaryredshift} shows the ratio of the odd- to even-parity trispectrum for an Euclid-like survey. We fix the shape and the viewing orientation of the quadrilateral and change the scale and redshift. The colouring on a log scaling is indicative of the amplitude of the corrections. As an overall trend, we can see that, for higher redshifts, the odd parity contributions become more dominant at lower scales. With upcoming surveys, such as MegaMapper, which will be probing $2 < z < 5$, it is interesting to see the effect of redshift on the odd parity part of the trispectrum. We will also be investigating this in more details in a follow-up paper.}

\section{Geometry setup} \label{app2}
\begin{figure}

\includegraphics[width=1\linewidth]{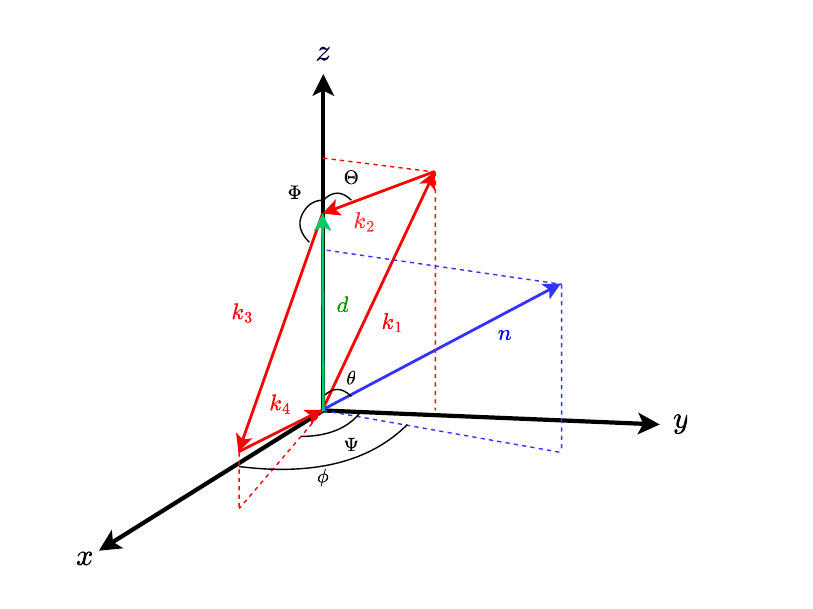}
\caption{Overview of the angles and vectors for the Fourier-space Trispectrum}
\label{fig:geom}
\end{figure}
Figure~\ref{fig:geom} shows the setup for the geometry of the trispectrum. The co-ordinate basis for the vectors for the quadrilateral are as follows - 
\begin{align}
    \boldsymbol{d} &= (0,0,d) \\ 
    \boldsymbol{n} &= (\sin \theta\cos\phi,\sin\theta\sin\phi,\cos\theta) \\ 
    \boldsymbol{k_1} &= d(0, s \sin\Theta, 1+s\cos\Theta )\\ 
    \boldsymbol{k_2} &= -d(0,s \sin\Theta, s \cos\Theta)\\ 
    \boldsymbol{k_3} &= d(t \sin\Phi \sin\Psi, t \sin\Phi \cos\Psi,  t\cos\Phi ) \\ 
    \boldsymbol{k_4} &= -d(t \sin\Phi \sin\Psi, t \sin\Phi \cos\Psi, t \cos\Phi + 1) 
\end{align}

We have fixed $\boldsymbol{d}$ along the z axis, and the triangle that $\boldsymbol{k}_1$ and $\boldsymbol{k}_2$ makes with $\boldsymbol{d}$ lies in the $y-z$ plane. We use $\theta,\phi$ to be the  angles which gives the orientation of the quadrilateral with respect to $\boldsymbol{n}$. $\Theta$ and $\Phi$ are the angles that vectors $\boldsymbol{k}_2$ and $\boldsymbol{k}_3$ make with the $z$ axis. $\Psi$ is the angle that $\boldsymbol{k}_4$ makes with the $y$ axis, which is therefore the folding angle of the quadrilateral. It determines how flat/folded the quadrilateral will be.  

\bibliographystyle{mnras}
\bibliography{ref}

\begin{thebibliography}{}
\makeatletter
\relax
\def\mn@urlcharsother{\let\do\@makeother \do\$\do\&\do\#\do\^\do\_\do\%\do\~}
\def\mn@doi{\begingroup\mn@urlcharsother \@ifnextchar [ {\mn@doi@} {\mn@doi@[]}}
\def\mn@doi@[#1]#2{\def\@tempa{#1}\ifx\@tempa\@empty \href {http://dx.doi.org/#2} {doi:#2}\else \href {http://dx.doi.org/#2} {#1}\fi \endgroup}
\def\mn@eprint#1#2{\mn@eprint@#1:#2::\@nil}
\def\mn@eprint@arXiv#1{\href {http://arxiv.org/abs/#1} {{\tt arXiv:#1}}}
\def\mn@eprint@dblp#1{\href {http://dblp.uni-trier.de/rec/bibtex/#1.xml} {dblp:#1}}
\def\mn@eprint@#1:#2:#3:#4\@nil{\def\@tempa {#1}\def\@tempb {#2}\def\@tempc {#3}\ifx \@tempc \@empty \let \@tempc \@tempb \let \@tempb \@tempa \fi \ifx \@tempb \@empty \def\@tempb {arXiv}\fi \@ifundefined {mn@eprint@\@tempb}{\@tempb:\@tempc}{\expandafter \expandafter \csname mn@eprint@\@tempb\endcsname \expandafter{\@tempc}}}

\bibitem[\protect\citeauthoryear{Abramo \& Bertacca}{Abramo \& Bertacca}{2017}]{Abramo:2017xnp}
Abramo L.~R.,  Bertacca D.,  2017, \mn@doi [Phys. Rev.] {10.1103/PhysRevD.96.123535}, D96, 123535

\bibitem[\protect\citeauthoryear{Bernardeau, Colombi, Gaztañaga  \& Scoccimarro}{Bernardeau et~al.}{2002}]{Bernardeau_2002}
Bernardeau F.,  Colombi S.,  Gaztañaga E.,   Scoccimarro R.,  2002, \mn@doi [Physics Reports] {10.1016/s0370-1573(02)00135-7}, 367, 1–248

\bibitem[\protect\citeauthoryear{Bertacca}{Bertacca}{2015}]{Bertacca:2014hwa}
Bertacca D.,  2015, \mn@doi [Class. Quant. Grav.] {10.1088/0264-9381/32/19/195011}, 32, 195011

\bibitem[\protect\citeauthoryear{Bertacca, Maartens, Raccanelli  \& Clarkson}{Bertacca et~al.}{2012}]{Bertacca:2012tp}
Bertacca D.,  Maartens R.,  Raccanelli A.,   Clarkson C.,  2012, \mn@doi [JCAP] {10.1088/1475-7516/2012/10/025}, 10, 025

\bibitem[\protect\citeauthoryear{Bertacca, Maartens  \& Clarkson}{Bertacca et~al.}{2014}]{Bertacca:2014dra}
Bertacca D.,  Maartens R.,   Clarkson C.,  2014, \mn@doi [JCAP] {10.1088/1475-7516/2014/09/037}, 1409, 037

\bibitem[\protect\citeauthoryear{Beutler et~al.,}{Beutler et~al.}{2014}]{Beutler_2014}
Beutler F.,  et~al., 2014, \mn@doi [MNRAS] {10.1093/mnras/stu1051}, 443, 1065

\bibitem[\protect\citeauthoryear{Bonvin}{Bonvin}{2014}]{Bonvin:2014owa}
Bonvin C.,  2014, \mn@doi [Class. Quant. Grav.] {10.1088/0264-9381/31/23/234002}, 31, 234002

\bibitem[\protect\citeauthoryear{Bonvin \& Durrer}{Bonvin \& Durrer}{2011}]{Bonvin:2011bg}
Bonvin C.,  Durrer R.,  2011, \mn@doi [Phys. Rev.] {10.1103/PhysRevD.84.063505}, D84, 063505

\bibitem[\protect\citeauthoryear{Bonvin, Hui  \& Gaztañaga}{Bonvin et~al.}{2014}]{Bonvin_2014}
Bonvin C.,  Hui L.,   Gaztañaga E.,  2014, \mn@doi [Phys. Rev.] {10.1103/physrevd.89.083535}, D89, 083535

\bibitem[\protect\citeauthoryear{Breton, Rasera, Taruya, Lacombe  \& Saga}{Breton et~al.}{2018}]{Breton_2018}
Breton M.-A.,  Rasera Y.,  Taruya A.,  Lacombe O.,   Saga S.,  2018, \mn@doi [MNRAS] {10.1093/mnras/sty3206}, 483, 2671

\bibitem[\protect\citeauthoryear{Cahn, Slepian  \& Hou}{Cahn et~al.}{2021}]{cahn2021test}
Cahn R.~N.,  Slepian Z.,   Hou J.,  2021 (\mn@eprint {arXiv} {2110.12004})

\bibitem[\protect\citeauthoryear{Camera, Fonseca, Maartens  \& Santos}{Camera et~al.}{2018}]{Camera:2018jys}
Camera S.,  Fonseca J.,  Maartens R.,   Santos M.~G.,  2018, \mn@doi [MNRAS] {10.1093/mnras/sty2284}, 481, 1251

\bibitem[\protect\citeauthoryear{{Cannon} et~al.,}{{Cannon} et~al.}{2006}]{2006MNRAS.372..425C}
{Cannon} R.,  et~al., 2006, \mn@doi [\mnras] {10.1111/j.1365-2966.2006.10875.x}, \href {https://ui.adsabs.harvard.edu/abs/2006MNRAS.372..425C} {372, 425}

\bibitem[\protect\citeauthoryear{Challinor \& Lewis}{Challinor \& Lewis}{2011}]{Challinor:2011bk}
Challinor A.,  Lewis A.,  2011, \mn@doi [Phys. Rev.] {10.1103/PhysRevD.84.043516}, D84, 043516

\bibitem[\protect\citeauthoryear{Clarkson, De~Weerd, Jolicoeur, Maartens  \& Umeh}{Clarkson et~al.}{2019}]{Clarkson:2018dwn}
Clarkson C.,  De~Weerd E.~M.,  Jolicoeur S.,  Maartens R.,   Umeh O.,  2019, \mn@doi [MNRAS] {10.1093/mnrasl/slz066}, 486, L101

\bibitem[\protect\citeauthoryear{De~Weerd, Clarkson, Jolicoeur, Maartens  \& Umeh}{De~Weerd et~al.}{2020}]{deWeerd:2019cae}
De~Weerd E.~M.,  Clarkson C.,  Jolicoeur S.,  Maartens R.,   Umeh O.,  2020, \mn@doi [JCAP] {10.1088/1475-7516/2020/05/018}, 05, 018

\bibitem[\protect\citeauthoryear{Desjacques, Jeong  \& Schmidt}{Desjacques et~al.}{2018}]{Desjacques:2016bnm}
Desjacques V.,  Jeong D.,   Schmidt F.,  2018, \mn@doi [Phys. Rept.] {10.1016/j.physrep.2017.12.002}, 733, 1

\bibitem[\protect\citeauthoryear{Di~Dio, Durrer, Marozzi  \& Montanari}{Di~Dio et~al.}{2014}]{DiDio:2014lka}
Di~Dio E.,  Durrer R.,  Marozzi G.,   Montanari F.,  2014, \mn@doi [JCAP] {10.1088/1475-7516/2014/12/017, 10.1088/1475-7516/2015/06/E01}, 12, 017

\bibitem[\protect\citeauthoryear{Dio \& Seljak}{Dio \& Seljak}{2019}]{Dio_2019}
Dio E.~D.,  Seljak U.,  2019, \mn@doi [Journal of Cosmology and Astroparticle Physics] {10.1088/1475-7516/2019/04/050}, 2019, 050

\bibitem[\protect\citeauthoryear{Eisenstein \& Hu}{Eisenstein \& Hu}{1998}]{Eisenstein:1997ik}
Eisenstein D.~J.,  Hu W.,  1998, \mn@doi [Astrophys. J.] {10.1086/305424}, 496, 605

\bibitem[\protect\citeauthoryear{Eskilt \& Komatsu}{Eskilt \& Komatsu}{2022}]{PhysRevD.106.063503}
Eskilt J.~R.,  Komatsu E.,  2022, \mn@doi [Phys. Rev.] {10.1103/PhysRevD.106.063503}, D106, 063503

\bibitem[\protect\citeauthoryear{Fonseca, Maartens  \& Santos}{Fonseca et~al.}{2018}]{Fonseca:2018hsu}
Fonseca J.,  Maartens R.,   Santos M.~G.,  2018, \mn@doi [MNRAS] {10.1093/mnras/sty1702}, 479, 3490

\bibitem[\protect\citeauthoryear{Giusarma, Alam, Zhu, Croft  \& Ho}{Giusarma et~al.}{2017}]{giusarma2017relativistic}
Giusarma E.,  Alam S.,  Zhu H.,  Croft R. A.~C.,   Ho S.,  2017 (\mn@eprint {arXiv} {1709.07854})

\bibitem[\protect\citeauthoryear{Hall \& Bonvin}{Hall \& Bonvin}{2017}]{Hall:2016bmm}
Hall A.,  Bonvin C.,  2017, \mn@doi [Phys. Rev.] {10.1103/PhysRevD.95.043530}, D95, 043530

\bibitem[\protect\citeauthoryear{Hou, Slepian  \& Cahn}{Hou et~al.}{2023}]{Hou_2023}
Hou J.,  Slepian Z.,   Cahn R.~N.,  2023, \mn@doi [MNRAS] {10.1093/mnras/stad1062}, 522, 5701

\bibitem[\protect\citeauthoryear{Jain \& Bertschinger}{Jain \& Bertschinger}{1994}]{Jain_1994}
Jain B.,  Bertschinger E.,  1994, \mn@doi [Astrophys. J.] {10.1086/174502}, 431, 495

\bibitem[\protect\citeauthoryear{Jeong \& Kamionkowski}{Jeong \& Kamionkowski}{2020}]{Jeong_2020}
Jeong D.,  Kamionkowski M.,  2020, \mn@doi [Phys. Rev. Lett.] {10.1103/physrevlett.124.041301}, 124, 041301

\bibitem[\protect\citeauthoryear{Jeong \& Schmidt}{Jeong \& Schmidt}{2020}]{Jeong_20201}
Jeong D.,  Schmidt F.,  2020, \mn@doi [Physical Review D] {10.1103/physrevd.102.023530}, 102

\bibitem[\protect\citeauthoryear{Jolicoeur, Umeh, Maartens  \& Clarkson}{Jolicoeur et~al.}{2017}]{Jolicoeur:2017nyt}
Jolicoeur S.,  Umeh O.,  Maartens R.,   Clarkson C.,  2017, \mn@doi [JCAP] {10.1088/1475-7516/2017/09/040}, 09, 040

\bibitem[\protect\citeauthoryear{Jolicoeur, Umeh, Maartens  \& Clarkson}{Jolicoeur et~al.}{2018}]{Jolicoeur:2017eyi}
Jolicoeur S.,  Umeh O.,  Maartens R.,   Clarkson C.,  2018, \mn@doi [JCAP] {10.1088/1475-7516/2018/03/036}, 03, 036

\bibitem[\protect\citeauthoryear{Jolicoeur, Maartens, De~Weerd, Umeh, Clarkson  \& Camera}{Jolicoeur et~al.}{2021}]{Jolicoeur:2020eup}
Jolicoeur S.,  Maartens R.,  De~Weerd E.~M.,  Umeh O.,  Clarkson C.,   Camera S.,  2021, \mn@doi [JCAP] {10.1088/1475-7516/2021/06/039}, 06, 039

\bibitem[\protect\citeauthoryear{Kaiser}{Kaiser}{1987}]{Kaiser:1987qv}
Kaiser N.,  1987, \mnras, 227, 1

\bibitem[\protect\citeauthoryear{Kamionkowski \& Souradeep}{Kamionkowski \& Souradeep}{2011}]{Kamionkowski_2011}
Kamionkowski M.,  Souradeep T.,  2011, \mn@doi [Phys. Rev.] {10.1103/physrevd.83.027301}, D83, 027301

\bibitem[\protect\citeauthoryear{Maartens, Jolicoeur, Umeh, De~Weerd, Clarkson  \& Camera}{Maartens et~al.}{2020}]{Maartens:2019yhx}
Maartens R.,  Jolicoeur S.,  Umeh O.,  De~Weerd E.~M.,  Clarkson C.,   Camera S.,  2020, \mn@doi [JCAP] {10.1088/1475-7516/2020/03/065}, 03, 065

\bibitem[\protect\citeauthoryear{{McDonald}}{{McDonald}}{2009}]{McDonald:2009dh}
{McDonald} P.,  2009, \mn@doi [JCAP] {10.1088/1475-7516/2009/11/026}, \href {https://ui.adsabs.harvard.edu/\#abs/2009JCAP...11..026M} {11, 026}

\bibitem[\protect\citeauthoryear{Minami \& Komatsu}{Minami \& Komatsu}{2020}]{Minami_2020}
Minami Y.,  Komatsu E.,  2020, \mn@doi [Phys. Rev. Lett.] {10.1103/physrevlett.125.221301}, 125, 221301

\bibitem[\protect\citeauthoryear{Nishizawa \& Kobayashi}{Nishizawa \& Kobayashi}{2018}]{Nishizawa_2018}
Nishizawa A.,  Kobayashi T.,  2018, \mn@doi [Phys. Rev.] {10.1103/physrevd.98.124018}, D98, 124018

\bibitem[\protect\citeauthoryear{Orlando, Pieroni  \& Ricciardone}{Orlando et~al.}{2021}]{Orlando_2021}
Orlando G.,  Pieroni M.,   Ricciardone A.,  2021, \mn@doi [JCAP] {10.1088/1475-7516/2021/03/069}, 03, 069

\bibitem[\protect\citeauthoryear{Philcox}{Philcox}{2022}]{Philcox_2022}
Philcox O.~H.,  2022, \mn@doi [Phys. Rev.] {10.1103/physrevd.106.063501}, D106, 063501

\bibitem[\protect\citeauthoryear{Raccanelli, Bertacca, Jeong, Neyrinck  \& Szalay}{Raccanelli et~al.}{2018}]{Raccanelli:2016avd}
Raccanelli A.,  Bertacca D.,  Jeong D.,  Neyrinck M.~C.,   Szalay A.~S.,  2018, \mn@doi [Phys. Dark Univ.] {10.1016/j.dark.2017.12.003}, 19, 109

\bibitem[\protect\citeauthoryear{Rodríguez-Torres et~al.,}{Rodríguez-Torres et~al.}{2016}]{Rodr_guez_Torres_2016}
Rodríguez-Torres S.~A.,  et~al., 2016, \mn@doi [MNRAS] {10.1093/mnras/stw1014}, 460, 1173

\bibitem[\protect\citeauthoryear{Scoccimarro, Zaldarriaga  \& Hui}{Scoccimarro et~al.}{1999}]{Scoccimarro_1999}
Scoccimarro R.,  Zaldarriaga M.,   Hui L.,  1999, \mn@doi [Astrophys. J.] {10.1086/308059}, 527, 1

\bibitem[\protect\citeauthoryear{Shiraishi, Nitta  \& Yokoyama}{Shiraishi et~al.}{2011}]{Shiraishi_2011}
Shiraishi M.,  Nitta D.,   Yokoyama S.,  2011, \mn@doi [Prog. Theor. Phys.] {10.1143/ptp.126.937}, 126, 937

\bibitem[\protect\citeauthoryear{Slepian \& Eisenstein}{Slepian \& Eisenstein}{2015}]{slepian2015new}
Slepian Z.,  Eisenstein D.~J.,  2015 (\mn@eprint {arXiv} {1510.04809})

\bibitem[\protect\citeauthoryear{Tellarini, Ross, Tasinato  \& Wands}{Tellarini et~al.}{2016}]{Tellarini:2016sgp}
Tellarini M.,  Ross A.~J.,  Tasinato G.,   Wands D.,  2016, \mn@doi [JCAP] {10.1088/1475-7516/2016/06/014}, 06, 014

\bibitem[\protect\citeauthoryear{Tokunaga, Stoeffler, Auguste, Shelkovnikov, Daussy, Amy-Klein, Chardonnet  \& Darquié}{Tokunaga et~al.}{2013}]{Tokunaga_2013}
Tokunaga S.~K.,  Stoeffler C.,  Auguste F.,  Shelkovnikov A.,  Daussy C.,  Amy-Klein A.,  Chardonnet C.,   Darquié B.,  2013, \mn@doi [Molecular Physics] {10.1080/00268976.2013.821186}, 111, 2363–2373

\bibitem[\protect\citeauthoryear{Umeh, Maartens  \& Santos}{Umeh et~al.}{2016}]{Umeh:2015gza}
Umeh O.,  Maartens R.,   Santos M.,  2016, \mn@doi [JCAP] {10.1088/1475-7516/2016/03/061}, 03, 061

\bibitem[\protect\citeauthoryear{Umeh, Jolicoeur, Maartens  \& Clarkson}{Umeh et~al.}{2017}]{Umeh:2016nuh}
Umeh O.,  Jolicoeur S.,  Maartens R.,   Clarkson C.,  2017, \mn@doi [JCAP] {10.1088/1475-7516/2017/03/034}, 03, 034

\bibitem[\protect\citeauthoryear{Verde, Heavens, Matarrese  \& Moscardini}{Verde et~al.}{1998}]{Verde:1998zr}
Verde L.,  Heavens A.~F.,  Matarrese S.,   Moscardini L.,  1998, \mn@doi [MNRAS] {10.1046/j.1365-8711.1998.01937.x}, 300, 747

\bibitem[\protect\citeauthoryear{Yankelevich \& Porciani}{Yankelevich \& Porciani}{2019}]{Yankelevich:2018uaz}
Yankelevich V.,  Porciani C.,  2019, \mn@doi [MNRAS] {10.1093/mnras/sty3143}, 483, 2078

\bibitem[\protect\citeauthoryear{Yoo}{Yoo}{2010}]{Yoo_2010}
Yoo J.,  2010, \mn@doi [Phys. Rev.] {10.1103/physrevd.82.083508}, D82, 083508

\bibitem[\protect\citeauthoryear{Yoo \& Zaldarriaga}{Yoo \& Zaldarriaga}{2014}]{Yoo:2014sfa}
Yoo J.,  Zaldarriaga M.,  2014, \mn@doi [Phys. Rev.] {10.1103/PhysRevD.90.023513}, D90, 023513

\makeatother
\end{thebibliography}

\end{document}